\documentstyle[11pt,newpasp,twoside,psfig]{article}
\def\etal{\emph{et al.\ }}
\def\mj{$M_{\rm J}$}
\def\mrat{$M_{\rm D}/M_*$}
\def\qmin{$Q_{\rm min}\ $}
\def\rd{$R_{\rm D}$}
\def\mdisk{$M_{\rm D}$}
\def\msun{$M_\odot$}

\begin{document}
\title{A Resolution Requirement for Disk Simulations Modeling
Collapse}
\author{Andrew F. Nelson}
\affil{School of Mathematics, University of Edinburgh,
Edinburgh Scotland, EH9 3JZ, UK}

\begin{abstract}
The accurate simulation of collapsed objects requires that a huge range
of spatial scales be well resolved if the result is not to be
contaminated by numerically induced fragmentation. In this context,
`insufficient resolution' means comparable to the local instability
wavelength in the system. I define a minimum resolution criterion for
simulations of disk systems, using the critical wavelength appropriate
for disks.
\end{abstract}

{\bf The Criteria in 2d and 3d}
\vskip 2mm

A condition on the minimum resolution to ensure the collapse in 3d is of
physical rather than numerical origin was defined by Truelove \etal
1997 using the ratio of the local grid resolution, $\delta x$, and
local Jeans wavelength, $\lambda_J$:
\begin{equation}\label{eq:True-cond}
\ \ \ \ \ \ \ \ \
J = {{\delta x}\over {\lambda_J}} \ \ \ \ \ {\rm where} \ \ \ \ \ 
\lambda_J = \left({{\pi c^2_s}\over{G\rho}}\right)^{1/2}.
\end{equation}
Here $c_s$ is the sound speed, $\rho$ is the volume density and $G$ is
the gravitational constant. An analogous condition can be obtained for
rotationally supported (i.e. disk) systems using the local Toomre
wavelength, $\lambda_T$, at which the disk is neutrally stable to
growing axisymmetric instabilities (see e.g. Lin \& Lau 1979): 
\begin{equation}\label{eq:Toom-cond}
T = {{\delta x}\over {\lambda_T}} \ \ \ \ \ {\rm where} \ \ \ \ \
\lambda_T =   {{2c_s^2 }\over{G\Sigma}}.
\end{equation}
Here again, $\delta x$ is the grid size, while $\Sigma$ is the local
surface density.

To obtain a form more useful in particle based numerical methods (e.g.
SPH), Bate \& Burkert (1997) using the Jeans mass to define an
analogous criterion for the maximum resolvable volume density. We can
obtain a similar quantity in 2d by defining a `Toomre mass' and a
maximum resolvable surface density. The two criteria are:
\begin{equation}\label{eq:max-dens}
\rho_{\rm max} =  \left({{2\pi^3}\over{9G^3}}\right)
{{c_s^6}\over{m_p N_{\rm reso}}} \ \ {\rm and} \ \ \
\Sigma_{\rm max} = \left({{\pi}\over{G^2}}\right)
                    {{c_s^4}\over{m_p N_{\rm reso}}}.
\end{equation}
$N_{\rm reso}$ is the number of particles (of mass $m_p$) required to
resolve the Jeans mass. 

In 3d, Truelove \etal 1997 found emperically that $J<1/4$ was
sufficient to suppress numerical collapse. Bate \& Burkert (1997)
arrived at a similar result for particle simulations: numerical
instability could be suppressed by resolving the local Jeans mass with
$N_{\rm reso}>2N_{\rm neigh}(\approx 100)$ neighbor particles. I will
define the same critical values for 2d as in 3d, i.e. $T<1/4$ and
$\sim100$ particles per Toomre mass, saving a more exact specification
for later work.

\vskip 2mm
{\bf Examples from the Literature}
\vskip 2mm

Nelson \etal (1998) modeled the evolution of self gravitating disks with
masses between 0.05 and 1.0 times the mass of the central star, using
SPH. From the suite of models in that work, I choose as an example the
model with disk dimensions \mrat=0.2 and \qmin=1.5, resolved with 7997
particles, implementing an isothermal equation of state. Using physical
dimensions, the disk has a radius \rd=50~AU, mass \mdisk$=0.2M_*$ and
$M_*=0.5$\msun. 

\begin{figure}\label{fig:DynI}
\begin{center}
\leavevmode\psfig{file=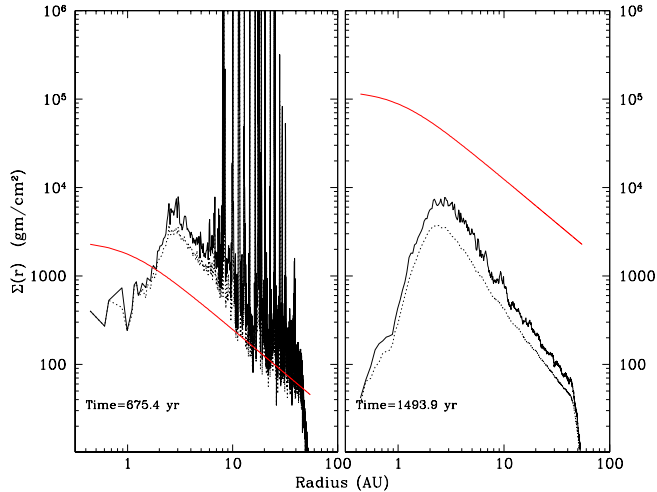,angle=-90,height=70mm,rheight=65mm}
\caption{Azimuth averaged (dotted line), maximum (solid) and critical
(red solid) surface densities for the Nelson \etal (1998) simulation
modeled with 8000 (left) and 397000 (right) particles}
\end{center}
\end{figure}
Figure 1 shows the surface densities produced at resolutions of 8000
(the original) and 397000 particles. At lower resolution, filamentary
spiral arms develop that later collapse into multiple clumps, as in the
Nelson \etal work, but such surface densities clearly violate criterion
of eq. \ref{eq:max-dens}. The higher resolution model did not
produce any clumps, even after evolving for three times as long (about
1500~yr). We originally concluded that although our physical model was
insufficent to correctly model clump formation, if it was to occur at
all it would be most likely between 10 and 40~AU. The statement
that the physical model in Nelson \etal was insufficient is certainly
valid (see e.g. Nelson \etal 2000). However, and clump formation was
definitely invalid: it was due to failure of the Toomre criterion and
not to any physical process, however modeled. 

\vskip 2mm

Armitage \& Hanson (1999) showed collapse occured in disks in which
a 1\mj\ point mass was embedded in a self-gravitating 0.1\msun\  disk
with an $r^{-3/2}$ density profile and no initial gap, resolved with
60000 SPH particles. The minimum stability of the disk using the Toomre
parameter, was \qmin=1.5, as in Nelson \etal. Temperature was defined
using a locally isothermal equation of state.

\begin{figure}
\begin{center}
\leavevmode\psfig{file=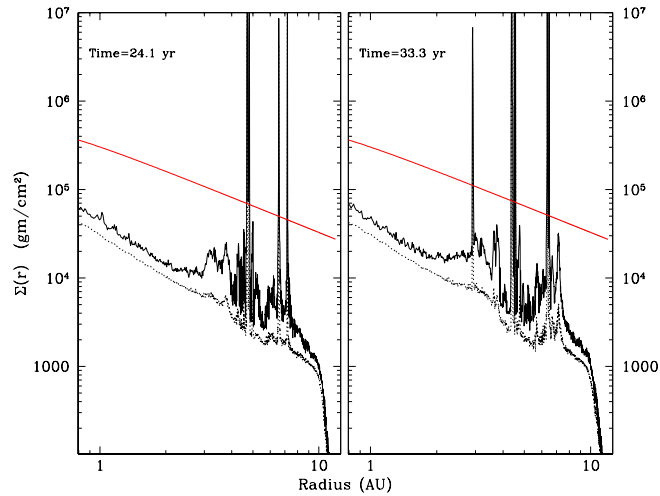,angle=-90,height=70mm,rheight=65mm}
\caption{Azimuth averaged (dotted line), maximum (solid) and critical
(red solid) surface densities for the Armitage \& Hanson (1999)
simulation.} \end{center} 
\end{figure}
Figure 2 shows a reproduction of this model. As in the original, the
planet both accretes a large amount of disk matter and generates large
amplitude spiral structures that began to collapse after $\sim20-25$yr.
Also as in the original, the initial radial positions correspond closely
to the outer and inner Lindblad resonances for the lowest order spiral
patterns, since the patterns reach their largest amplitudes there. These
collapsed objects are clearly numerical artifacts and the author's
conclusion that if a planet could grow sufficiently rapidly (to 4-5\mj)
sufficiently early enough in the disk's evolution, it would trigger
further planet formation in the disk, is not supported.

\vskip 2mm

In the most recent of a series of papers studying collapse in disks,
Boss (2002) models a disk in 3d (reflected through the midplane),
between 4 and 20~AU, with $100\times23\times256$ grid zones
asymmetrically distributed primarily near the disk midplane. He includes
an ideal gas equation of state and a radiative cooling model. He evolves
a disk for 359~yr, when the simulation begins to violate the Jeans
criterion due to the onset of clump formation. 

Figure 3 shows the critical wavelength at 359~yr defined using the Jeans
criterion (Boss: personal communication) and using eq.
\ref{eq:Toom-cond}. The values from the two criteria coincide at the
location of the clump. Elsewhere, the Toomre wavelength exceeds the
Jeans wavelength. Thus the evolution of the Boss simulation leading up
to its termination passes this test, and the conclusions made from it
are valid numerically.

\begin{figure}
\begin{center}
\leavevmode\psfig{file=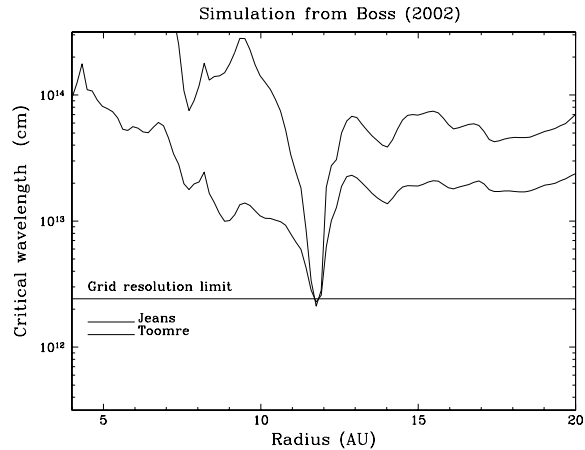,angle=-90,height=60mm,rheight=55mm}
\caption{The failure of the Jeans criterion (solid line) was used by Boss
(2002) to define the end of his simulation. The Toomre criterion
applicable for circumstellar disks (heavy solid line) shows a near
identical critical wavelength at this resolution.} 
\end{center}
\end{figure}

\vskip 2mm
{\bf Concluding Questions and Remaining Work}

{\bf What is the proportionality factor for the 2d criterion?} So far, I
have only assumed that the spatial resolution required in the 3d case
would carry over in 2d as well, but this may not be correct. 

{\bf Which critical wavelength criterion is more applicable: Toomre or
Jeans?} The universe is (spatially) a 3d place, so a criterion based on
only two coordinates is necessarily approximate. On the other hand, the
geometry of a disk is highly flattened. The assumption made in the Jeans
analysis--of an infinite (in all directions) homogenous medium--is not
satisfied and may incorrectly capture the physical behavior of the
system. Further, correspondence between the two criteria in the forming
clump of Boss's simulation may be coincidental. The two criteria have
differing functional dependences on the temperature.

\vskip 3mm

Except for the simulation discussed in figure 3, generously shared prior
to its publication by Alan Boss, the computations reported here were
performed using the UK Astrophysical Fluids Facility (UKAFF). I also
gratefully acknowledge financial support from UKAFF.

\vskip 1mm

\noindent
{\bf References}

\noindent
Armitage, P. J., Hansen, B. M. S., 1999, Nature, 402, 633 

\noindent
Bate, M. R., Burkert, A., 1997, MNRAS, 288, 1060

\noindent
Boss, A. P., 2002, ApJ, 576, In press 

\noindent
Lin, C. C. \& Lau, Y. Y., 1979, Studies in Applied
Mathematics, 60, 97

\noindent
Nelson, A. F., Benz, W., Adams, F. C., Arnett, W. D.,
1998, ApJ, 502, 342   

\noindent
Nelson, A. F., Benz, W., Ruzmaikina, T. V., 2000,
ApJ, 529, 357

\noindent
Truelove, J. K., Klein, R. I., McKee, C. F., 
Holliman, J., H., Howell, L. H., Greenough, J. A., 1997, ApJL, 
489, 179

\end{document}